\PassOptionsToPackage{unicode}{hyperref}
\PassOptionsToPackage{hyphens}{url}
\PassOptionsToPackage{dvipsnames,svgnames,x11names}{xcolor}
\documentclass[
  12pt]{article}

\usepackage{amsmath,amssymb}
\usepackage{iftex}
\ifPDFTeX
  \usepackage[T1]{fontenc}
  \usepackage[utf8]{inputenc}
  \usepackage{textcomp} 
\else 
  \usepackage{unicode-math}
  \defaultfontfeatures{Scale=MatchLowercase}
  \defaultfontfeatures[\rmfamily]{Ligatures=TeX,Scale=1}
\fi
\usepackage{lmodern}
\ifPDFTeX\else  
\fi
\IfFileExists{upquote.sty}{\usepackage{upquote}}{}
\IfFileExists{microtype.sty}{
  \usepackage[]{microtype}
  \UseMicrotypeSet[protrusion]{basicmath} 
}{}
\makeatletter
\@ifundefined{KOMAClassName}{
  \IfFileExists{parskip.sty}{%
    \usepackage{parskip}
  }{
    \setlength{\parindent}{0pt}
    \setlength{\parskip}{6pt plus 2pt minus 1pt}}
}{
  \KOMAoptions{parskip=half}}
\makeatother
\usepackage{xcolor}
\makeatletter
\ifx\paragraph\undefined\else
  \let\oldparagraph\paragraph
  \renewcommand{\paragraph}{
    \@ifstar
      \xxxParagraphStar
      \xxxParagraphNoStar
  }
  \newcommand{\xxxParagraphStar}[1]{\oldparagraph*{#1}\mbox{}}
  \newcommand{\xxxParagraphNoStar}[1]{\oldparagraph{#1}\mbox{}}
\fi
\ifx\subparagraph\undefined\else
  \let\oldsubparagraph\subparagraph
  \renewcommand{\subparagraph}{
    \@ifstar
      \xxxSubParagraphStar
      \xxxSubParagraphNoStar
  }
  \newcommand{\xxxSubParagraphStar}[1]{\oldsubparagraph*{#1}\mbox{}}
  \newcommand{\xxxSubParagraphNoStar}[1]{\oldsubparagraph{#1}\mbox{}}
\fi
\makeatother

\usepackage{longtable,booktabs,array}
\usepackage{calc} 
\usepackage{etoolbox}
\makeatletter
\patchcmd\longtable{\par}{\if@noskipsec\mbox{}\fi\par}{}{}
\makeatother
\IfFileExists{footnotehyper.sty}{\usepackage{footnotehyper}}{\usepackage{footnote}}
\makesavenoteenv{longtable}
\usepackage{graphicx}
\makeatletter
\def\maxwidth{\ifdim\Gin@nat@width>\linewidth\linewidth\else\Gin@nat@width\fi}
\def\maxheight{\ifdim\Gin@nat@height>\textheight\textheight\else\Gin@nat@height\fi}
\makeatother
\setkeys{Gin}{width=\maxwidth,height=\maxheight,keepaspectratio}
\makeatletter
\def\fps@figure{htbp}
\makeatother

\makeatletter
\@ifpackageloaded{caption}{}{\usepackage{caption}}
\AtBeginDocument{%
\ifdefined\contentsname
  \renewcommand*\contentsname{Table of contents}
\else
  \newcommand\contentsname{Table of contents}
\fi
\ifdefined\listfigurename
  \renewcommand*\listfigurename{List of Figures}
\else
  \newcommand\listfigurename{List of Figures}
\fi
\ifdefined\listtablename
  \renewcommand*\listtablename{List of Tables}
\else
  \newcommand\listtablename{List of Tables}
\fi
\ifdefined\figurename
  \renewcommand*\figurename{Figure}
\else
  \newcommand\figurename{Figure}
\fi
\ifdefined\tablename
  \renewcommand*\tablename{Table}
\else
  \newcommand\tablename{Table}
\fi
}
\@ifpackageloaded{float}{}{\usepackage{float}}
\floatstyle{ruled}
\@ifundefined{c@chapter}{\newfloat{codelisting}{h}{lop}}{\newfloat{codelisting}{h}{lop}[chapter]}
\floatname{codelisting}{Listing}

\makeatother
\makeatletter
\@ifpackageloaded{caption}{}{\usepackage{caption}}
\@ifpackageloaded{subcaption}{}{\usepackage{subcaption}}
\makeatother

\ifLuaTeX
  \usepackage{selnolig}  
\fi
\usepackage[]{natbib}
\usepackage{bookmark}

\IfFileExists{xurl.sty}{\usepackage{xurl}}{} 
\hypersetup{
  pdftitle={Bayesian Measures of Leverage and Influence},
  colorlinks=true,
  linkcolor={blue},
  filecolor={Maroon},
  citecolor={Blue},
  urlcolor={Blue},
  pdfcreator={LaTeX via pandoc}}

\newcommand{\btheta}{\boldsymbol{\theta}}
\newcommand{\epsvec}{\boldsymbol{\epsilon}}
\newcommand{\deltavec}{\boldsymbol{\delta}}
\newcommand{\bY}{\mathbf Y}

\newcommand{\bx}{\mathbf x}
\newcommand{\Yvec}{\mathbf{Y}}
\newcommand{\wvec}{\boldsymbol{\omega}}
\newcommand{\rvec}{\mathbf r}
\newcommand{\bone}{\mathbf 1}
\newcommand{\pW}{p_W}
\newcommand{\pD}{p_D}
\newcommand{\pV}{p_V}
\newcommand{\rE}{\text{E}}

\newcommand{\anon}{1}

\begin{document}

\def\spacingset#1{\renewcommand{\baselinestretch}%
{#1}\small\normalsize} \spacingset{1}


\if1\anon
{
  \title{\bf Bayesian Measures of Leverage and Influence}
  \author{Martyn Plummer\thanks{}\hspace{.2cm}\\
    Department of Statistics, University of Warwick}
  \maketitle
} \fi

\if0\anon
{
  \bigskip
  \bigskip
  \bigskip
  \begin{center}
    {\LARGE\bf Bayesian Measures of Leverage and Influence}
\end{center}
  \medskip
} \fi

\bigskip
\begin{abstract}
Leverage and influence are regression diagnostics that are used to measure the sensitivity of parameter estimates to changes in the data. In this article, Bayesian leverage and influence diagnostics are derived from a local sensitivity framework in which the case weights of individual observations are perturbed.  The resulting diagnostics can be applied to any Bayesian model and are easy to estimate using Markov Chain Monte Carlo.

Bayesian measures of leverage and influence are closely related to predictive information criteria that are commonly used for Bayesian model choice. The penalty terms for these information criteria can be reinterpreted in terms of local sensitivity diagnostics. This connection helps to understand differences between the various information criteria that have been proposed in the literature.

A comparison between leverage and influence measures leads to a new diagnostic for outlier detection. By considering multivariate case weight perturbations, groups of observations can be highlighted that are collectively outliers. This may suggest ways to improve the model. 
A diagnostic for sensitivity to the learning rate is also proposed that may be interpreted as a measure of prior-data conflict. This diagnostic can be adapted to measure cross-conflict between different parts of the data.
\end{abstract}

\noindent%
\vfill

\newpage


\section{Introduction}

Leverage and influence are two closely related regression model diagnostics. Influence is the effect of deleting an observation on the fit of the model or, more generally, the effect of perturbing its case weight. Leverage is the sensitivity of fitted values to changes in the outcome. High leverage observations show an unusual pattern of predictor variables compared with the other observations and may also be influential. 

In frequentist inference for linear models, Cook's statistic is widely used to assess influence \citep{CookWeisberg1982}. Leverage in linear models is measured by hat-values, which are the diagonal elements of the hat matrix that projects observed values onto fitted values. Approximate versions of these diagnostics can be calculated for generalized linear models (GLMs) via the iterative weighted least squares algorithm, which represents the fitted GLM as a weighted linear model. Leverage diagnostics have been extended to linear mixed models \citep{Demidenko2005, Nobre2011}. Influence measures based on case deletion have also been extended to linear mixed models \citep{Christensen1992}, but a more versatile approach to frequentist influence analysis has been the local influence approach of \citet{Cook1986} that perturbs the case weights of each observation. This has led to the development of local influence diagnostics for linear mixed models \citep{Beckman1987, Lesaffre1998}, generalized linear mixed models \citep{Ouwens2001, Rakhmawati2017}, and models with missing data \citep{ZhuLee2001}.

In this article, a local sensitivity approach is used to characterize Bayesian influence and leverage by measuring the effect of case weight perturbations on the posterior distribution. This reveals a surprising connection with predictive information criteria, which are used to assess goodness of fit in Bayesian modeling \citep{Gelman2014}. The title of this article is a self-conscious reference to {\em Bayesian Measures of Model Complexity and Fit}, the paper that introduced the deviance information criterion (DIC) \citep{Spiegelhalter2002DIC}. Since the DIC was introduced, various other information criteria have been proposed \citep{Gelman2004BDA, Plummer2008, Watanabe2010}. All of these combine a measure of model fit with a complexity penalty. It will be shown that these penalties can be reinterpreted in terms of local sensitivity diagnostics.

The rest of the article proceeds as follows: Section~\ref{section:notation} introduces notation that will be used throughout the article and a simple linear model that can be used to illustrate the diagnostics. Section~\ref{section:influence} discusses Bayesian influence diagnostics and their relation to the Widely Applicable Information Criterion (WAIC) \citep{Watanabe2009Book, Watanabe2010}. Section~\ref{section:leverage} discusses Bayesian leverage diagnostics and their relation to DIC. Section~\ref{section:multivariate_perturbations} discusses multivariate perturbations. Section~\ref{section:outlier_detection} introduces a proposal to detect outlying observations by comparing leverage and influence diagnostics. Section~\ref{section:examples} illustrates these ideas with selection of real-world datasets. In section~\ref{section:prior_data_conflict}, the complexity penalty  proposed by \citet{Gelman2004BDA} is examined as a diagnostic for prior-data conflict and this is illustrated with further real-world data. Section~\ref{section:discussion} ends with a discussion.

\section{Notation}
\label{section:notation}

Suppose that the data consist of pairs $(\bx_i, Y_i)$ for $i = 1 \ldots n$, where the outcome variables $Y_1, \ldots Y_n$ are conditionally independent given the predictor variables $\bx_1, \ldots \bx_n$. Consider a probability model for $Y_i$ parameterized by $\btheta \in \Theta \subseteq \mathbb{R}^k$. 
A commonly used framework for Bayesian local influence is the weighted pseudo-likelihood
\begin{equation}
\label{eq:weighted_likelihood}
    L_{\wvec}(\btheta, \bY) = \prod_{i=1}^n p(Y_i \mid \bx_i, \btheta)^{w_i}
\end{equation}
where $w_1 \ldots w_n$ are non-negative real-valued case weights. If $w_i$ is an integer, then the weighted likelihood (\ref{eq:weighted_likelihood}) corresponds to observing $w_i$ identical copies of observation $i$. This includes $w_i = 0$, which corresponds to no copies or, equivalently, deleting the observation.

If $\pi(\btheta)$ is the prior density function of $\btheta$ then the pseudo-posterior $p_{\wvec}(\btheta \mid \Yvec)$ is derived by applying Bayes theorem using the weighted pseudo-likelihood
\begin{equation*}
p_{\wvec}(\btheta \mid \Yvec) \propto \pi(\btheta) L_{\wvec}(\btheta, \bY)
\end{equation*}
This reduces to the usual posterior $p(\btheta \mid \bY)$ when $\wvec = \bone$. For notational convenience, the dependency on predictor variables $\bx_1 \ldots \bx_n$ is implicit. This convention is used below whenever there is no ambiguity.

The effect of using perturbed case weights $\wvec \neq \bone$ on the posterior can be measured using a phi-divergence between $p(\btheta \mid \Yvec)$ and $p_{\wvec}(\btheta \mid \Yvec)$.
\begin{equation*}
\Delta(\wvec) = \int \varphi\left( \frac{p_{\wvec}(\btheta \mid \bY)}{p(\btheta \mid \bY)} \right) p(\btheta \mid \bY) d\btheta 
\end{equation*}
where $\varphi: \mathbb{R}^+_0 \to \mathbb{R}$ is a convex function such that $\varphi(1) = 0$. The family of phi-divergences includes the Kullback-Leibler divergence (for  $\varphi(r) = - \log(r)$), the reverse Kullback-Leibler divergence (for $\varphi = r \log(r)$),the squared Hellinger distance (for $\varphi(r) =  1 - \sqrt{r}$), and the total variation distance (for $\varphi(r) = \left| r - 1 \right|/2$). In this article, attention will be restricted to divergences where $\varphi(r)$ is twice differentiable at $r=1$.

\subsection{Normal linear model}
\label{section:linear}

None of the diagnostics discussed in this article depend on simplifying assumptions such as linearity. Nevertheless, it can be helpful to illustrate how measures of leverage and influence behave in simple models. This will be done with the normal-normal linear model.
\begin{equation}
\label{eq:linear_model}
\begin{aligned}
Y_i & \sim N(\bx_i^T\btheta, \,\sigma^2) \\
\btheta & \sim N_k({\mathbf 0}, \, \Psi^{-1}) 
\end{aligned}
\end{equation}
where the variance $\sigma^2$ is fixed. The non-informative limiting prior is obtained as the largest eigenvalue of the prior precision $\Psi$ tends to zero. Under this limit, the posterior expectation $\overline{\btheta} = \rE(\btheta \mid \bY)$ coincides with the maximum likelihood estimate $\widehat{\btheta}$. 

Let $X$ be the design matrix, an $n \times p$ matrix such that row $i$ is $\bx^T_i$. Then the fitted values for $\Yvec$ are defined by $\overline{Y} = X \overline{\btheta} = H \Yvec$, where $H$ is the {\em hat matrix}
\[
H = X^T (\Psi \sigma^2 + X^T  X)^{-1} X 
\]
In the non-informative limit $\overline{\bY} \to \widehat{\bY} = X \widehat{\btheta}$. The diagonal elements of the hat matrix $h_{11} \ldots h_{nn}$ are the hat-values which are the frequentist measures of leverage for each observation since $\partial \widehat{Y}_i/\partial Y_i = h_{ii}$. 

\section{Local Influence}
\label{section:influence}

If all case weights are close to 1 then, assuming sufficient regularity conditions, the divergence $\Delta(\wvec)$ can be approximated by a Taylor series expansion
\begin{equation}
\label{eq:influence_taylor_vector}
\Delta(\wvec) = \frac{\varphi''(1)}{2} (\wvec - \bone)^T V (\wvec - \bone) + o\left(\left\|\wvec - \bone\right\|^2\right)
\end{equation}
where $V$ is the posterior variance-covariance matrix of the log-likelihood contributions: 
\begin{equation}
    V_{ij} = \text{Cov}_{\btheta} \left[ \log p(Y_i \mid \btheta), \log p(Y_j \mid \btheta) \mid \bY \right]
    \label{eq:V}
\end{equation}
Thus all phi-divergences have the same local behaviour up to a constant of proportionality $\varphi''(1)$. When only observation $i$ is perturbed ({\em i.e.} $w_j = 1$ for $j \neq i$), the divergence is
\begin{equation}
\Delta(w_i) \approx \frac{\varphi''(1)}{2} \; (w_i - 1)^2 V_{ii}
\label{eq:influence_taylor_scalar}
\end{equation}
Based on this approximation, \citet{MillarStewart2007} proposed $V_{ii} = \text{Var}_{\btheta} \left( \log p (Y_i \mid \btheta) \mid \Yvec \right)$ as a Bayesian influence diagnostic. Further theoretical elaboration of this idea was provided by \citet{vanderLinde2007}.  Let
\[
\text{LINF}_i = V_{ii} = \text{Var}_{\btheta} \left[ \log p (Y_i \mid \btheta) \mid \Yvec \right]
\]
be the local influence of observation $i$. 

\subsection{Local influence and WAIC}

The Widely Applicable Information Criterion (WAIC) \cite{Watanabe2010} is a predictive information criterion derived from singular learning theory \citep{Watanabe2009Book}. In fact, there are two versions of WAIC which use different loss functions to assess the goodness of fit of a Bayesian model. The first version is based on the Bayes generalization error.
\begin{equation*}
BL_g =  \rE_{\bx^*Y^*} \left\{ - \log \rE_{\btheta} \left[ p(Y^* \mid \btheta, \bx^*)  
\mid \bY \right] \right\}
\end{equation*}
where $(\bx^*, Y^*)$ is a sample from the true distribution of the data. Since this distribution is unknown, we replace the expectation with an empirical average over the data to get the Bayes training loss.
\begin{equation*}
BL_t = - \frac{1}{n} \sum_{i=1}^n \log \rE_{\btheta} \left[ p(Y_i \mid \btheta, \bx_i) \mid \Yvec \right]
\end{equation*}
The second version of WAIC is derived from the Gibbs generalization error
\begin{equation*}
GL_g = \rE_{\bx^*Y^*} \left\{ - \rE_{\btheta} \left[ \log p(Y^* \mid \btheta, \bx^* ) \mid \Yvec \right]
\right\}
\end{equation*}
and its empirical estimate, the Gibbs training loss: 
\begin{equation*}
GL_t = - 
\frac{1}{n} \sum_{i=1}^n \rE_{\btheta} \left[ \log p(Y_i \mid \btheta, \bx_i)
\mid
\Yvec \right]
\end{equation*}
In both cases, the training loss systematically underestimates the generalization error because it uses the data $\bY$ twice: once to estimate the parameters $\btheta$ and again to estimate the expectation of the loss. This bias can be corrected by adding a penalty to give Widely Applicable Information Criteria
\begin{align*}
\text{WAIC}_1 & = BL_t + \frac{\pW}{n} \\
\text{WAIC}_2 & = GL_t + \frac{\pW}{n}
\end{align*}
where the WAIC penalty is the same in both cases:
\begin{equation*}
\pW = \sum_{i=1}^n  \text{Var}_{\btheta} \left[ \log p (Y_i \mid \btheta, \bx_i) \mid \Yvec \right]
\end{equation*}
In singular learning theory, $\pW/2$ is an estimate of the  {\em singular fluctuation}, one of two birational invariants that define the dimensions of a statistical model \citep{Watanabe2009Book}. The other is the {\em real log canonical threshold}, which is used in the construction of the Bayesian Information Criterion (BIC) for singular models \citep{Watanabe2013,DrtonPlummer2017}.
Both $\text{WAIC}_1$ and $\text{WAIC}_2$ are asymptotically unbiased estimates of their respective generalization errors to order $o(n^{-1})$ even when the model is misspecified \citep{Watanabe2010}.

The WAIC penalty $\pW$ can be rewritten as the sum of the local influence values
\[
\pW = \sum_{i=1}^n V_{ii} = \sum_{i=1}^n \text{LINF}_i
\]
Thus observations that are more influential contribute more to the WAIC penalty \citep{Millar2018}.

\citet{Watanabe2009Book} also proposed an alternate penalty based on the difference between the Gibbs and Bayes training losses:
\[
    \pW^{*} =  2n (GL_t - BL_t)
\]
This alternative WAIC penalty can also be interpreted in terms of sensitivity. If the quadratic approximation (\ref{eq:influence_taylor_scalar}) holds in the interval $w_i \in [0,2]$ then we can, paradoxically, approximate the effect of deleting an observation $(w_i = 0)$ by doubling its weight instead $(w_i = 2)$. 
This is represented by the weight vector $\wvec = \bone + \deltavec^i$, where $\delta^i_j = I(i=j)$. This not a local perturbation, so different phi-divergences will give different expressions. Choosing the Kullback-Leibler divergence gives
\[
\Delta_{KL} (
\mathbf{\bone} + \deltavec^i) = - \rE_{\btheta} \left[ \log p(Y_i \mid \btheta) \mid \Yvec \right] +
\log \rE_{\btheta} \left[ p(Y_i \mid \btheta) \mid \Yvec \right]
\]
which is the difference between the contributions of observation $i$ to the Gibbs training loss and the Bayes training loss. Hence, if we define the {\em doubling influence} for observation $i$ as
\[
\text{DINF}_i = 2 \Delta_{KL} (\mathbf{\bone} + \deltavec^i)
\]
then the alternate WAIC penalty can be written as
\[
\pW^{*} = \sum_{i=1}^n \text{DINF}_i
\]

The two WAIC penalties $\pW$ and $\pW^{*}$ are asymptotically equivalent \citep{Watanabe2009Book}, but they can behave differently in small samples \citep{Gelman2014}.

\subsection{Influence in the linear model}

For the linear model of section~\ref{section:linear}, the local influence of observation $i$ can be expressed in terms of its residual $r_i = y_i - \overline{y}_i$ and hat-value $h_{ii}$. We can do the same for the doubling influence $\text{DINF}_i$, as well as the zeroing influence $\text{ZINF}_i = \Delta_{KL}(\bone - \deltavec^i)$ which sets the weight of observation $i$ to zero, equivalent to deleting the observation.
\begin{align*}
\text{LINF}_i & = \frac{r^2_i h_{ii}}{\sigma^2} + \frac{h^2_{ii}}{2} \\
\text{DINF}_i & = \frac{r^2_i h_{ii}}{\sigma^2 (1 + h_{ii})} + h_{ii} - \log(1 + h_{ii}) \\
\text{ZINF}_i & = \frac{r^2_i h_{ii}}{\sigma^2 (1 - h_{ii})} - h_{ii} - \log(1 - h_{ii})
\end{align*}
For large $n$, these influence measures are all equivalent as $h_{ii} = O(n^{-1})$. In small samples they may be quite different for observations with large hat-values. In particular there is a strict ordering $\text{DINF}_i < \text{LINF}_i < \text{ZINF}_i$ for $h_{ii} \in (0, 1]$. If the aim is to estimate the influence of deleting an observation (ZINF) then LINF is a better approximation than DINF. This is consistent with the finding of \citet{Gelman2014} that $\pW$ gives a better approximation than $\pW^{*}$ to cross-validation loss.

In the non-informative limit, all three Bayesian influence measures may be compared with Cook's distance \citep{CookWeisberg1982}
\[
D^{\text{Cook}}_i = \frac{1}{k} \frac{r^2_i h_{ii}}{\sigma^2(1-h_{ii})^2} 
\]
For high-leverage observations, as the hat-value $h_{ii}$ approaches its maximum value of 1, $\text{ZINF}_i$ and $D^{\text{Cook}}_i$, which are both based on case deletion, tend to infinity whereas $\text{LINF}_i$ and $\text{DINF}_i$ do not.

\section{Leverage}
\label{section:leverage}

In linear models the frequentist leverage of an observation is a function of the predictor variables $\bx_i$ and does not depend on the observed outcome $Y_i$. The same holds approximately in more complex models, although the leverage of observation $i$ may depend indirectly on $Y_i$ via its contribution to parameter estimation. In Bayesian inference,  independence from the outcome can be attained by adopting a posterior predictive approach. Suppose we have replicate observations $\bY^r = (Y^r_1, \ldots Y^r_n)$ and the extra information from these replicates is incorporated using a weighted pseudo-likelihood
\[
p_{\wvec^r} (\btheta \mid \bY, \bY^r)  \propto \pi(\btheta) \, L_{\bone}(\btheta, \bY)  \, L_{\wvec^r}(\btheta, \bY^r)
\]
This reduces to the usual posterior $p(\btheta \mid \bY)$ when $\wvec^r = {\mathbf 0}$.

Now suppose that $\bY^r$ is sampled from the posterior predictive distribution given $\bY$. The expected divergence $\overline{\Delta}(\wvec^r)$ between $p(\btheta \mid \bY)$ and $p_{\wvec^r} (\btheta \mid \bY, \bY^r)$ is
\begin{equation}
\label{eq:expected_local_divergence}
\overline{\Delta}(\wvec^r)  = \iint \varphi \left( 
\frac{p_{\wvec^r}(\btheta \mid \bY, \bY^r)}{p(\btheta \mid \bY)} 
\right) 
p(\btheta, \bY^r \mid \bY)\, d\bY^r \, d\btheta
\end{equation}

Assuming sufficient regularity conditions, a Taylor expansion for values of $\wvec^r$ close to zero gives
\begin{equation}
\overline{\Delta}(\wvec^r) = \varphi'(1) \sum_{i=1}^n w^r_i \, h_i + o\left(\left\|\wvec^r\right\|\right)
\label{eq:leverage_taylor_vector}
\end{equation}
where 
\begin{equation}
\label{eq:hat_values}
h_i = \rE_{Y_i^r} \left\{ \rE_{\btheta} \left[ \log p(Y_i^r \mid \btheta) \mid \bY, Y_i^r \right] - \rE_{\btheta} \left[ \log p(Y_i^r \mid \btheta) \mid \bY \right]  \mid \bY \right\}
\end{equation}
As with local influence, all phi-divergences have the same local behavior up to a constant of proportionality. In contrast to equation (\ref{eq:influence_taylor_vector}), which is quadratic in terms of the case weights for the observations $w_1 \ldots w_n$, equation (\ref{eq:hat_values}) is linear in the replicate case weights $w^r_1 \ldots w^r_n$.

If only one replicate observation is introduced ({\em i.e.} $w^r_j = 0$ for $j \neq i$), the expected local divergence is
\[
\overline{\Delta}(w^r_i) \approx \varphi'(1) w^r_i h_i
\]
This suggests that $h_i$ can be used as a measure of leverage for observation $i$, and can be considered the Bayesian hat-value.
\[
\text{LLEV}_i = h_i
\]

The double expectation on the right hand side of equation (\ref{eq:hat_values}) can be expressed in kernel form \citep{Plummer2008}
\begin{equation}
h_i = h(\bx_i) = \iint \Delta^{r}_{KL}
\left(
\btheta^{(1)},
\btheta^{(2)}
\mid
\bx_i
\right) \,
p(\btheta^{(1)} \mid \bY) \, 
p(\btheta^{(2)} \mid \bY) 
\, d\btheta^{(1)} \, d\btheta^{(2)}
\label{eq:hat_values_kernel}
\end{equation}

The integrand in (\ref{eq:hat_values_kernel}) is the Kullback-Leibler divergence between the predictive distributions for a replicate outcome $Y^r$ evaluated at two different values of the parameter $\btheta$.
\begin{equation}
\Delta^{r}_{KL}(\btheta^{(1)}, \btheta^{(2)} \mid \bx)  = 
\rE_{Y^r} \left[
\log \left(
\frac{p(Y^r \mid \btheta^{(1)}, \bx)} 
     {p(Y^r \mid \btheta^{(2)}, \bx)}
\right)\,
\middle\vert \,
\bx, \btheta^{(1)}
\right]
\label{eq:KL_for_hat_value}
\end{equation}
Using MCMC, (\ref{eq:hat_values_kernel}) can be estimated by drawing independent samples $\btheta^{(1)}, \btheta^{(2)}$ from the posterior distribution using parallel chains.  If $\Delta_{KL}$ is not available in closed form then it can be replaced with an unbiased empirical estimate.

The Bayesian hat-value $h(\bx)$ is not limited to the observed values $\bx_1, \ldots \bx_n$. In machine learning terms, it can be calculated for test data as well as the training data. A machine learning model deployed in an open-world scenario may be exposed to test data that differ substantially from the training data. Such out-of-distribution data might come from a class that was not included in the training data, a phenomenon known as semantic shift. The need to correctly handle out-of-distribution test data has led to the development of machine learning methods that are conceptually close to outlier detection such as anomaly detection, novelty detection, and open set recognition \citep{Salehi2022, Yang2024}. 
For a Bayesian classifier, the hat-value $h(\bx)$ might be a useful way to detect semantic shift. The kernel form (\ref{eq:hat_values_kernel},\ref{eq:KL_for_hat_value}) shows that a high value of $h(\bx)$ is associated with inconsistent predictions, in the sense that the posterior gives support to different values of $\btheta$ with very different predictive distributions for $Y$. This may be interpreted as a lack of confidence in the marginal prediction of $Y$. We might consider setting a cutoff value for $h(\bx)$ above which the classifier should simply refuse to return a prediction.

\subsection{Local leverage and DIC}
\label{section:dic}

\citet{Spiegelhalter2002DIC} introduced the Deviance Information Criterion (DIC) as a Bayesian generalization of the Akaike Information Criterion \citep{Akaike1973}. Let $D(\btheta) = -2 \log p(\Yvec \mid \btheta)$ be the deviance. Then
\[
\text{DIC} = D(\overline{\btheta}) + 2 \, \pD
\]
DIC combines a measure of model fit or adequacy $D(\overline{\btheta})$ with a complexity penalty 
\begin{equation*}
\pD = D(\overline{\btheta}) - \overline{D}
\end{equation*}
where $\overline{D} = \rE_{\btheta}( D(\btheta) \mid \Yvec)$ is the posterior expectation of $D$. Other variants of DIC are possible using different plug-in estimates of $\btheta$ such as the posterior median or mode.

\citet{Spiegelhalter2002DIC} called $\pD$ the {\em effective number of parameters}, noting that if there is a non-informative limit for the prior distribution $\pi(\btheta)$ then $\pD \to k$, the dimension of the parameter space $\Theta$.
The penalty $\pD$ has some theoretical problems which were the subject of vigorous discussion when the DIC was introduced \citep{Brooks2002discussion}. Firstly, $\pD$ is not parametrization invariant. Secondly, $\pD$ may be negative when the likelihood is not log-concave in $\btheta$. To address these problems, \citet{Plummer2008} suggested an alternative definition of the effective number of parameters 
\begin{equation}
\label{eq:pDstar}
p_{D}^{*} = \sum_{i=1}^n \rE_{Y_i^r} \left[ \rE_{\btheta} \left( \log p(Y_i^r \mid \btheta) \mid \bY, Y_i^r \right) - \rE_{\btheta} \left( \log p(Y_i^r \mid \btheta) \mid \bY \right)  \mid \bY \right]
\end{equation}
The derivation of $\pD^{*}$ is based on an explicit attempt to measure the optimism of the expected log likelihood. Equation (\ref{eq:pDstar}) compares the expected log likelihood of a new observation $Y^r_i$ under two posteriors: one that only uses the current observations $\Yvec$, and one that incorporates information from $Y^r_i$ into the posterior of $\btheta$. The difference between these two expectations is an approximation to the rational penalty that should be paid for using the observation $Y_i$ twice \citep{Plummer2008}.

The right hand side of equation (\ref{eq:pDstar}) is the sum of the Bayesian hat-values from equation (\ref{eq:hat_values}). Hence we can write
\[
\pD^{*} = \sum_{i=1}^m h_i = \sum_{i=1}^n \text{LLEV}_i
\]
In the same way that $\pW$ and $\pW^{*}$ can be expressed as the sum of influence values, $\pD^{*}$ can be expressed as the sum of local leverage values.

Local leverage  has an interesting connection with Value-of-Information (VoI) analysis, a branch of decision theory that is concerned with the cost-effectiveness of collecting additional data. \citet{Jackson2022} discuss how VoI can assess the sensitivity of models to different sources of uncertainty and also how it can be applied to estimation problems. Suppose that the utility of the posterior distribution $p(\btheta \mid \Yvec)$ is measured by the expected log-likelihood of a replicate data set $\rE_{\btheta} \left[ \log p(\Yvec^r \mid \btheta ) \mid \bY\right]$. The Bayesian hat-value $h_i$ is the expected increase in utility from observing $Y^r_i$ before calculating the posterior distribution. In VoI terms this is called the {\em expected value of sample information} (EVSI). Likewise, $\pD^* = \sum_i h_i$ is the EVSI for observing the whole replicate data set $\Yvec^r$.

\subsection{Leverage in the linear model}

In the linear model of section~\ref{section:linear}, the Bayesian hat-value of observation $i$ is the $i$th diagonal element of the hat matrix and hence coincides with the frequentist measure of leverage, {\em i.e.} $h_i = h_{ii}$. Additionally,
$\pD = \pD^* = \text{tr}\left(H\right)$.
The connection between $\pD$ and the trace of the hat matrix in the linear model was highlighted by \citet{Spiegelhalter2002DIC}, who also noted that there had been previous proposals to characterize the effective number of parameters as the trace of the hat matrix. 
They also suggested that individual contributions to $\pD$ could be used as leverages. The derivation of the Bayesian hat-value from the expected local divergence (\ref{eq:expected_local_divergence}) formalizes this relationship and justifies its application to general Bayesian models.

\section{Multivariate perturbations}
\label{section:multivariate_perturbations}

While $\text{LLEV}_i$ and $\text{LINF}_i$ may be useful ways to characterize the leverage and influence of individual observations, these diagnostics are derived from a much richer set of model perturbations in equations (\ref{eq:influence_taylor_vector}) and (\ref{eq:leverage_taylor_vector}). More useful diagnostic information may be obtained by considering this larger perturbation set. This presents a problem that the local influence and leverage perturbations are not directly comparable. The case weights $\wvec$ allow both positive and negative perturbations, whereas only positive perturbations of the replicate weights $\wvec^r$ are allowed. Moreover, the divergence used to measure influence is locally quadratic in $\wvec - \bone$ but the expected divergence used to measure leverage is locally linear in $\wvec^r$. In order to make the local behavior of these divergences comparable, we consider perturbations of the form
\begin{equation}
\begin{split}
w_i & = 1 + \epsilon_i \\
w^r_i & = \epsilon^2_i
\end{split}
\label{eq:multivariate_perturbations}
\end{equation}
so that both divergences are locally quadratic in $\epsvec$. This is a redundant parametrization of the possible perturbations of the replicate weights, but it can be a useful way to construct a set of complementary perturbations. If $\epsvec^{1} \ldots \epsvec^{n}$ is an orthonormal basis of $\mathbb{R}^n$ then the corresponding replicate weights $\wvec^{r1} \ldots \wvec^{rn}$ represent a factorization of the likelihood for the replicate observations
\[
L_{\bone}(\btheta, \bY^r) = 
\prod_{j=1}^n L_{\wvec^{rj}}(\btheta, \bY^r)
\]
and $\sum_i w^{rj}_i = 1$ so that each factor holds the equivalent information of one observation.

Two key developments in frequentist inference provide a model for multivariate perturbations in Bayesian inference. Firstly, \citet{Cook1986} developed local influence diagnostics based on the likelihood displacement
\[
LD(\wvec) = 2 \left[ \log L(\widehat{\btheta}, \bY) - \log L(\widehat{\btheta}_{\wvec}, \bY)\right]
\]
where $\widehat{\btheta}_{\wvec}$ is the estimate of $\btheta$ obtained by maximizing the weighted pseudo-likelihood $L_{\wvec}(\btheta, \bY)$. For case weights close to 1, the likelihood displacement is
\[
LD(\wvec) = (\wvec - \bone)^T M (\wvec - \bone)/2 + o\left(\left\|\wvec - \bone\right\|^2\right)
\]
where $M$ is the Hessian matrix of $LD(\wvec)$. Eigenvalue decomposition of $M$ indicates the multivariate perturbations of the case weights that correspond to maximum influence. 

\citet{Cook1986} suggested that this approach could be adapted to Bayesian inference by substituting the Kullback-Leibler divergence for the the likelihood displacement and this idea was taken up by \citet{McCulloch1989}. Similarly, \citet{Lavine1992} looked at multivariate perturbations of case-weights but looked at divergence of the posterior predictive distribution $p(\Yvec^r \mid \Yvec)$ instead of the poster distribution of the parameters. 
The observation by \cite{MillarStewart2007} that the Hessian matrix of the posterior divergence with respect to the case weights is the variance-covariance matrix $V$ (\ref{eq:V}) cuts through the algebraic complexity of this approach and enables straightforward estimation using MCMC. The principal eigenvector of $V$ indicates the maximally influential perturbation. More generally, multivariate analysis techniques applied to $V$ can reveal clusters of influential points that may otherwise be hidden by masking and swamping effects \citep{Thomas2018}.

The second key development in frequentist inference was the proposal of \citet{Poon1999} to summarize the influence of a perturbation $\epsvec$ in terms of the conformal normal curvature.
\[
\frac{\epsvec^T M \epsvec}{\text{tr}(M^2)^{1/2} (\epsvec^T \epsvec)}
\]
which depends on the direction but not the magnitude of the perturbation, and is invariant under conformal transformations of the perturbation space. A simplified definition of conformal normal curvature suitable for positive definite matrices was given by \citet{ZhuLee2001}
\[
\frac{\epsvec^T M \epsvec}{\text{tr}(M) (\epsvec^T \epsvec)}
\]
Using this simpler definition, define the conformal local influence
\[
\text{CLINF}(\epsvec)  = \frac{\epsvec^T V \epsvec}{\text{tr}(V) (\epsvec^T\epsvec)} = \frac{\sum_i \sum_j V_{ij} \epsilon_i \epsilon_j}{(\sum_i v_{ii}) (\sum_j \epsilon^2_j)}
\]
and conformal local leverage
\[
\text{CLLEV}(\epsvec)  = \frac{\epsvec^T H \epsvec}{\text{tr}(H) (\epsvec^T\epsvec)} = \frac{\sum_i h_i \epsilon^2_i}{(\sum_i h_i) (\sum_j \epsilon^2_j)}
\]
where $H= \text{diag}(h_1 \ldots h_n)$.
If $\epsvec^1 \ldots \epsvec^n$ form an orthogonal basis for $\mathbb{R}^n$ then
\[
\sum_i \text{CLINF}(\epsvec^i) = \sum_i \text{CLLEV}(\epsvec^i) = 1
\]
Hence the conformal local influence/leverage measures the proportion of total influence/leverage attributable to a perturbation in direction $\epsvec^i$. Univariate perturbations give conformal local influence and leverage statistics for each observation
\begin{align*}
\text{CLINF}_i & = \text{CLINF}(\deltavec^i) = \frac{V_{ii}}{\sum_j V_{jj}} = 
\frac{\text{LINF}_i}{\pW}\\
\text{CLLEV}_i & = \text{CLLEV}(\deltavec^i) = \frac{h_{i}}{\sum_j h_{j}} =
\frac{\text{LLEV}_i}{\pD^*}
\end{align*}
which are the proportional contributions of observation $i$ to $\pW$ and $\pD^*$, respectively.


\section{Outlier detection}
\label{section:outlier_detection}

Separating the concepts of leverage and influence allows a more fine-grained analysis of anomalous observations. If an observation has high leverage then it may be considered influential by design and a closer look at its predictor variables $\bx_i$ is required. If an observation has high influence but low leverage then its outcome $Y_i$ may have an unusual or extreme value, in which case it would be classified as an outlier.  This idea was formalized by \citet{Parsons2022}. Working in a VoI framework, they proposed an outlier diagnostic based on the ratio between a measure of influence and a measure of leverage, both based on case deletion. Adapting this idea to local perturbations,
define the conformal local outlier statistic as the ratio between conformal local influence and leverage.
\[
\text{CLOUT}_i = \frac{\text{CLINF}_i}{\text{CLLEV}_i}
= \frac{V_{ii}}{h_i} \frac{\sum_j h_j}{\sum_j V_{jj}}
\]
Observations with high values of $\text{CLOUT}_i$ may be considered outliers. The CLOUT diagnostic is non-directional, which has both advantages and disadvantages. For scalar outcomes it does not tell us if $Y_i$ is unusually high or low but it does generalize easily to multivariate or nominal outcomes.



Generalizing the CLOUT diagnostic to multivariate perturbations of the form (\ref{eq:multivariate_perturbations}), define the outlier matrix. 
\begin{equation}
\Omega = \frac{\text{tr}(H)}{\text{tr}(V)} H^{-1/2} V H^{-1/2}
\label{eq:outlier_matrix}
\end{equation}
and the generalized CLOUT diagnostic
\[
\text{CLOUT}(\epsvec) = \frac{\epsvec^T \Omega \epsvec}{\epsvec^T \epsvec}
\]
The diagonal elements of $\Omega$ then correspond to univariate perturbations of individual observations, {\em i.e.} $\text{CLOUT}(\deltavec^i) = \Omega_{ii} = \text{CLOUT}_i$.

Eigenvalue decomposition of the outlier matrix $\Omega$ indicates which multivariate perturbations are informative about outlying observations. If $\lambda_1 \geq \lambda_2 \geq \ldots \geq \lambda_n$ are the eigenvalues and $\epsvec^1 \ldots \epsvec^n$ are the corresponding eigenvectors of $\Omega$ then $\text{CLOUT}(\epsvec^i) = \lambda_i$. Hence the principal eigenvector $\epsvec^1$ is the perturbation that is maximally outlying and the observations that contribute to this perturbation can be identified by inspecting the elements of $\epsvec^1$.

Focusing on the principal eigenvalue works best when $\Omega$ has a large spectral gap so that the principal eigenvector accounts for most of the anomalous observations. In high-dimensional problems it is possible for $\Omega$ to have multiple large eigenvalues that all correspond to perturbations of interest. One approach to this situation is to use clustering methods to identify groups of observations \citep{Thomas2018}. Another possibility, which will be illustrated in section~\ref{section:bikesharing}, is to look at the aggregate effect of multiple perturbations. This can be done by noting that the univariate CLOUT statistics can be written
\[
\text{CLOUT}_i = \Omega_{ii} = \sum_{j=1}^n \lambda_j (\epsilon^j_i)^2
\]
Hence we can define a truncated CLOUT statistic that only considers contributions from the $m$ largest eigenvalues
\[
\text{CLOUT}^{(m)}_i\ = \sum_{j=1}^m \lambda_j \left(\epsilon^j_i\right)^2
\]

\section{Examples}
\label{section:examples}

\subsection{Abalone data}

The data set Abalone \citep{Abalone94} contains data from a marine survey of the edible marine shellfish abalone. The data set contains a small number of anomalous observations which can be readily found using exploratory data analysis. The aim of this example is to show how these can be found after model fitting using local sensitivity diagnostics.

\begin{figure}
\centering
\resizebox{\textwidth}{!}{\includegraphics{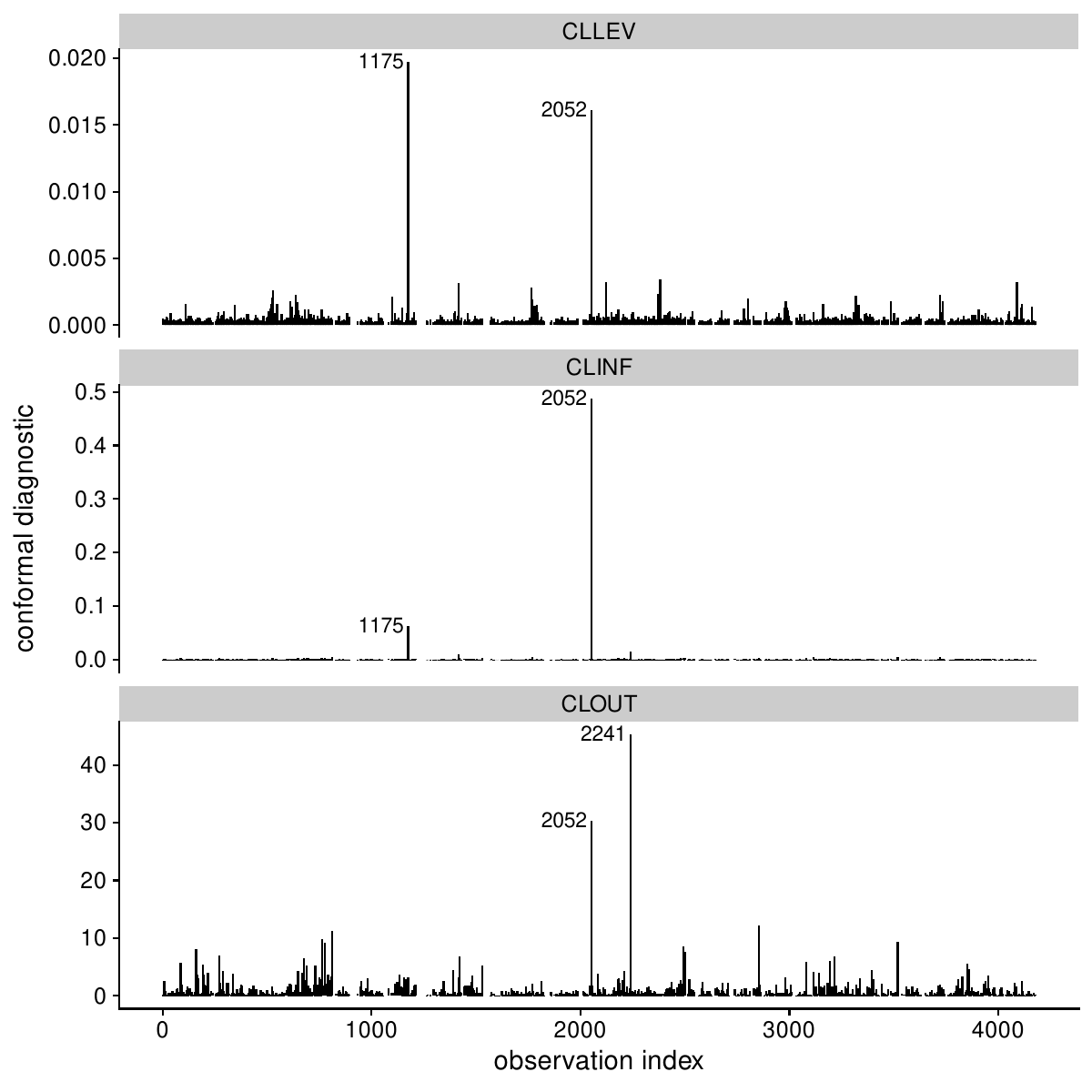}}
\caption{\label{fig:abalone-influence} Conformal local leverage (CLLEV), influence (CLINF), and outlyingness (CLOUT) for the Abalone data.} 
\end{figure}

Consider the problem of predicting the amount of meat (the ``shucked weight'') obtained from an abalone based on length, diameter, height, weight, and sex. After removing the immature individuals, there are n=2835 abalone in the data set. A gamma GLM was fitted with log link. 
Figure~\ref{fig:abalone-influence} shows conformal diagnostics for each observation. Observations 1175 and 2052 have unusually high leverage (CLLEV). Both are also highly influential (CLINF) with 2052 accounting for nearly half of the conformal influence. The CLOUT statistic reveals observation 2241 as an outlier, although it does not have high leverage or influence.

Table~\ref{tab:abalone} shows data for the three anomalous observations. High-leverage observations 1175 and 2052 have unusual height compared with their other dimensions: 1175 is very flat and 2052 is very tall. These values are so far outside the range of variation in the rest of the data that they are most likely data errors. Outlying observation 2241 has unusually low shucked weight. In fact it has the lowest ratio of shucked weight to whole weight in the whole data set. 

\begin{table}[tbp]
\centering
\begin{tabular}{llrrrrr}
\hline
row     & sex & length & diameter & height & whole & shucked \\ 
number  &     &        &          &        & weight & weight \\
\hline
1175 & F & 127 & 99 & {\bf 3} & 231.3 & 102.3 \\
2052 & F & 91 & 71 & {\bf 226} & 118.8 & 66.4 \\
2241 & M & 83 & 63 & 25 & 77.6 & {\bf 13.6} \\
\hline
\end{tabular}
\caption{\label{tab:abalone} Data for the anomalous observations in the Abalone data set found by the diagnostic plots in figure~\ref{fig:abalone-influence}. Unusual
values are highlighted in bold.}
\end{table}

\subsection{Bike sharing data}
\label{section:bikesharing}

The data set Bike Sharing \citep{BikeSharing2013} contains hourly data on the number of bicycles rented out by the company Capital Bikeshare DC in the years 2011 and 2012. This is a useful data set for illustrating outlier analysis because an explanation for anomalous observations can be found in public media sources.

The number of bicycles hired by users without a subscription -- known as ``casual users'' --  was modeled using Poisson regression with  predictors based on time of day, season, weather, and indicator variables for weekends and public holidays. The training data was restricted to the year 2011.

\begin{figure}[tbp]
\centering
\resizebox{\textwidth}{!}{\includegraphics{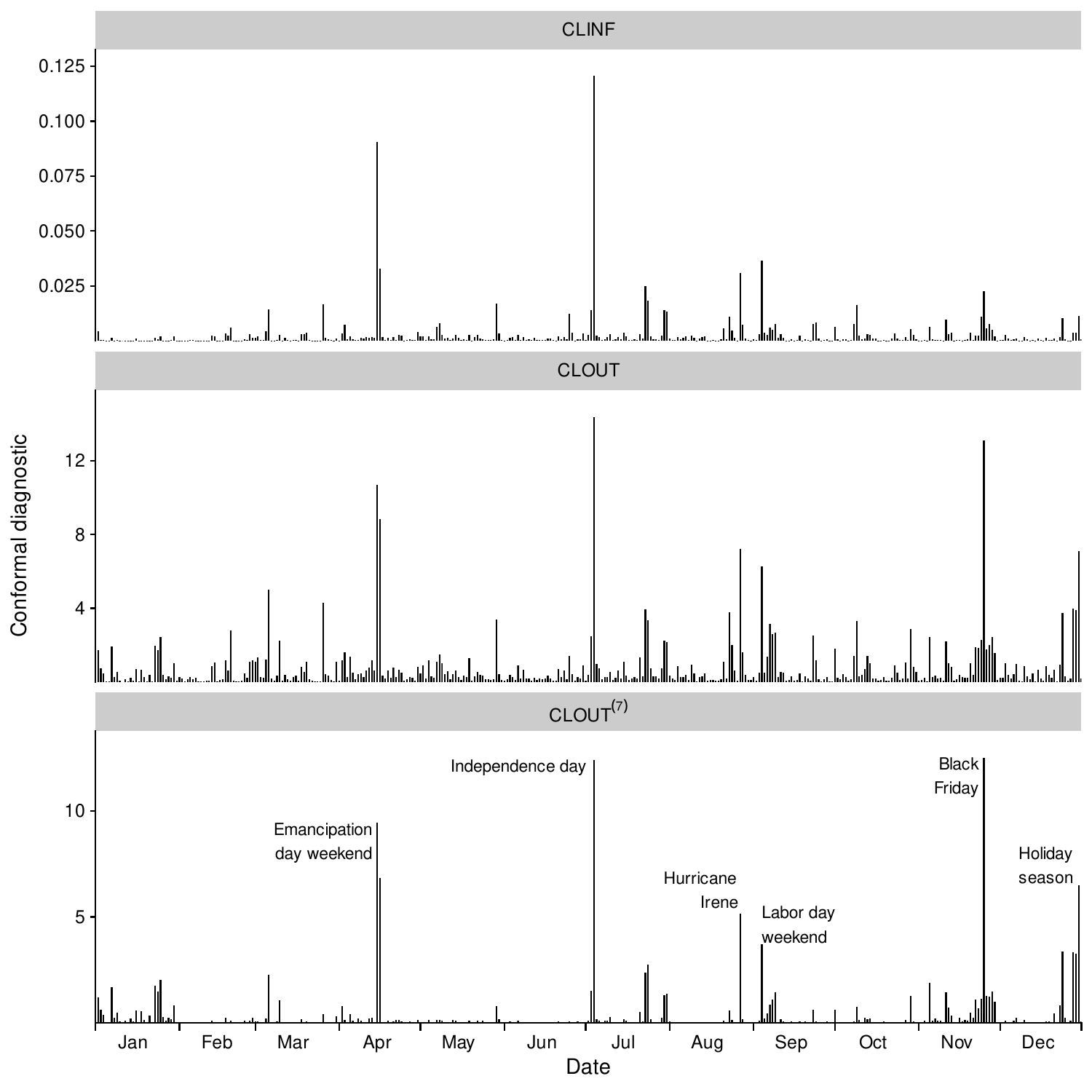}}
\caption{\label{fig:bikeshare-influence} Diagnostics for the bike sharing data. Top panel: Conformal local influence. Middle panel: conformal local outliers. Bottom panel: outliers based on the first 7 principal components of the outlier matrix.} 
\end{figure}

The model is based on hourly rental data, with a separate outcome for each hour of the day giving $n=8760$ independent observations. For the purposes of evaluating local sensitivity, we are not constrained to keep this fine-grained view of the data but can aggregate observations into whole days. The top panel of Figure~\ref{fig:bikeshare-influence} shows CLINF for each day of 2011. There are two outstandingly influential periods which are identified below. There were no days with outstandingly CLLEV so a leverage plot is not shown. The middle panel shows potential outliers based on CLOUT. A cleaner version of the outlier plot is obtained by truncating the eigenvalue expansion of the outlier matrix $\Omega$, as described in section~\ref{section:multivariate_perturbations}. A scree plot (not shown) suggests that the first 7 principal components of $\Omega$ may be sufficient to capture the most interesting variation and so the bottom panel of figure~\ref{fig:bikeshare-influence} shows $\text{CLOUT}^{(7)}$ Most important outliers are associated with public holidays including the days before ({\em e.g.} the Sunday before Labor Day) or after ({\em e.g.} Black Friday after Thanksgiving).  The CLOUT statistic also highlights 27 August 2011, when hurricane Irene hit the North Eastern United States, as an outlier.

\begin{figure}[tbp]
\centering
\resizebox{\textwidth}{!}{\includegraphics{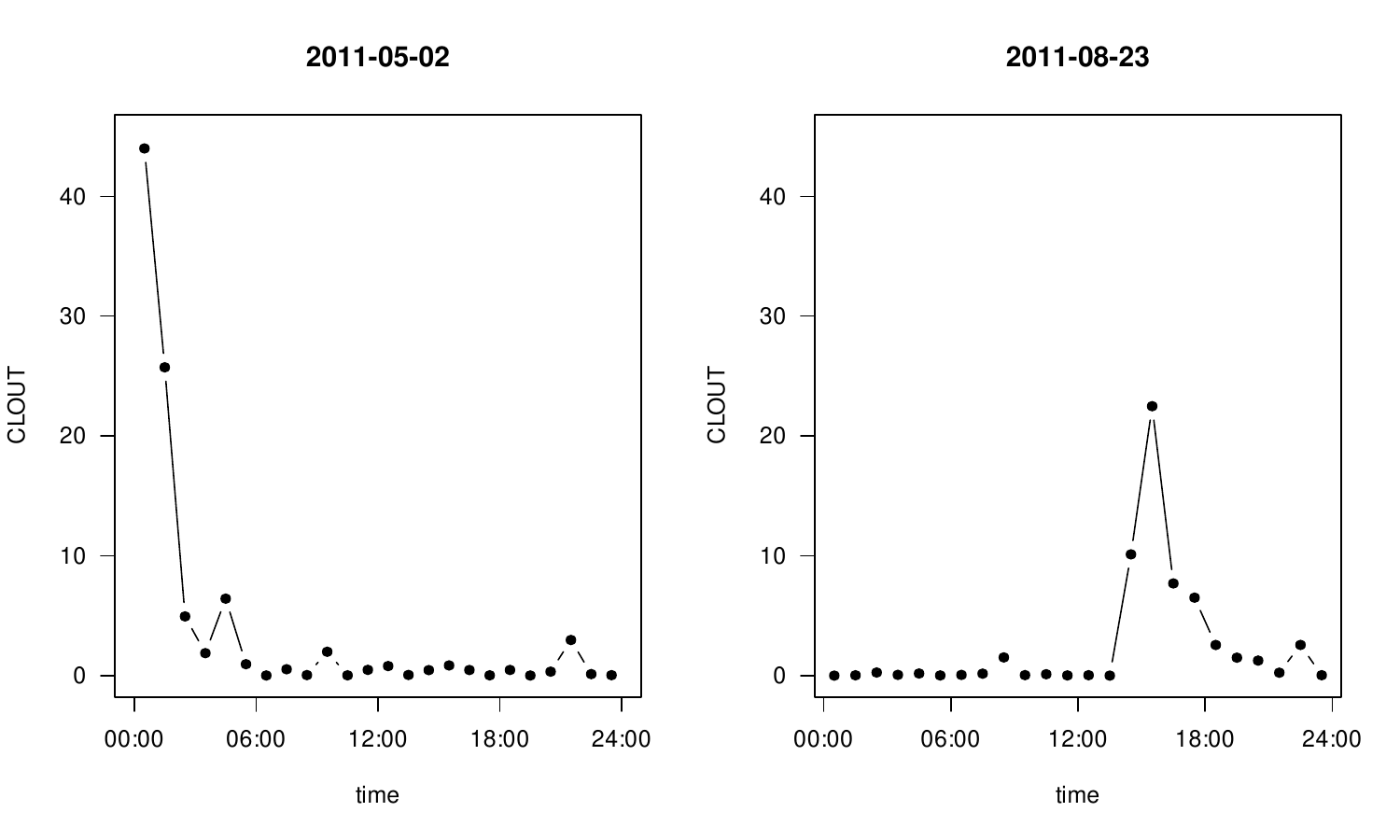}}
\caption{\label{fig:bikeshare-hourly} Hourly CLOUT statistics for the bike sharing data on two days when normal usage was disrupted by unusual events.} 
\end{figure}

The analysis of daily influence and leverage captures important differences in user behavior that extend across whole days, but misses short term changes. These can be detected by analyzing the data at its original hourly scale. The most outlying hour, as measured by the $\text{CLOUT}$ statistic was from 0:00 to 1:00 on 2 May 2011. Excluding this date and previously identified holiday periods, the second most outlying hour was 15:00 to 16:00 on 23 August. Figure~\ref{fig:bikeshare-hourly} shows hourly CLOUT statistics for these two days.  On 1 May 2011 at 23:35, President Obama announced the death of Osama Bin Laden on live television. In the following hours, there were spontaneous gatherings in the center of Washington, leading to an increase in the number of bicycles hired in the small hours of the morning. On 26 August at 13:51, there was a magnitude 5.8 earthquake in the Piedmont region of Virginia. This severely disrupted transport networks in Washington DC, leading to an increase in bicycle hiring as commuters sought alternative forms of transport. 

\section{Prior-data conflict}
\label{section:prior_data_conflict}

\citet[p.~182]{Gelman2004BDA} suggested an alternative penalty for DIC \[
\pV = 2 \,\text{Var}_{\btheta} \left( \sum_{i=1}^n \log p(Y_i \mid \btheta) \;\middle|\; \bY \right)
\]
This superficially resembles the WAIC penalty $\pW$. However $\pV$ is based on the variance of the sum of the log-likelihood contributions from each observations, whereas $\pW$ is the sum of the variances. Moreover $\pV$ is scaled by a factor of 2 to ensure $\pV \to k$ in the non-informative limit. In a retrospective review of DIC, \cite{Spiegelhalter2014} reported that they had considered $\pV$ as an alternative penalty to $\pD$, but rejected it.

Like all of the previously considered penalty functions $\pV$ can be interpreted in terms of local influence. From equation (\ref{eq:influence_taylor_vector}), $\pV$ is the local influence associated with the perturbation $\wvec = (1 + \epsilon) \bone$ for small values of $\epsilon$. The conformal local influence of this perturbation is
\[
\text{CLINF}(\bone) = \frac{\bone^T V \bone}{n \, \text{tr}(V)} = \frac{1}{2n}\frac{\pV}{\pW}
\]
Since both $\pV$ and $\pW$ are calibrated to be equal to $k$ for non-informative priors, the ratio $\pV/\pW$ can be used directly as a measure of local influence with a reference value of 1. Note that the conformal local leverage of this perturbation is constant: $\text{CLLEV}(\bone) = 1/n$.

The perturbation $\wvec = (1 + \epsilon) \bone$ corresponds to an extension of Bayesian inference in which the likelihood is raised to a power $\tau = 1 + \epsilon$ called the temperature or learning rate. Using a tempered likelihood with $\tau \in (0,1]$ reduces the contribution of the likelihood to the posterior distribution. This has been suggested as a way to improve robustness to model misspecification \citep{GrunwaldVanOmmen2017,Bhattacharya2019} and can be interpreted as conditioning on a coarsened version of the data \citep{MillerDunson2019}. The ratio $\pV/\pW$ thus measures sensitivity to perturbations of the learning rate $\tau$. Since $\tau$ controls the balance of information between the prior and the likelihood, a large ratio $\pV/\pW$ can be interpreted as indicating prior-data conflict.

\subsection{Prior-data conflict in the linear model}

Returning to the linear model of section \ref{section:linear}, the two penalties can be written in terms of residuals $r_i$ and the hat matrix $H$
\begin{align}
\pW & = \sum_{i=1}^n 
\left( \frac{h_{ii} r_i^2}{\sigma^2} + \frac{h^2_{ii}}{2} \right) 
        \label{eq:pW_linear} \\
\pV & = 2 \left( \frac{\rvec^T H \rvec}{\sigma^2} + \frac{\text{tr}(H^2)}{2} \right)
\label{eq:pV_linear}
\end{align}
Equation (\ref{eq:pV_linear}) is a multivariate version of (\ref{eq:pW_linear}), with an additional calibration factor of 2. Despite this algebraic similarity, the two expressions behave quite differently in the non-informative limit. In this case, the first term in (\ref{eq:pW_linear}) is $k + O_p(n^{-1/2})$ and the second term is $O(n^{-1})$. Conversely, the first term in (\ref{eq:pV_linear}) vanishes for a non-informative prior, and the second term is $k$. Thus, although both $\pW$ and $\pV$ tend to $k$ for a non-informative prior, they do so in different ways. This suggests that they are measuring different underlying quantities. The first term in (\ref{eq:pV_linear}) can be rewritten as
\[
2 \frac{\rvec^T H \rvec}{\sigma^2} = \frac{2}{\sigma^2} 
\left(\widehat{\bY} - \overline{\bY}\right)^T H \left(\widehat{\bY} - \overline{\bY}\right)
\]
Hence $\pV$ grows when an informative prior pulls the Bayesian fitted values $\overline{\bY}$ away from the frequentist fitted values $\widehat{\bY}$ or, equivalently, pulls $\overline{\btheta}$ away from $\widehat{\btheta}$.

\subsection{Examples of prior-data conflict}

Two examples below illustrate the use of $\pV/\pW$ as a diagnostic for prior-data conflict. Both examples use the binomial distribution $Y_i \sim \text{Bin}(\pi_i, m_i)$ where $\pi_i = \pi_i(\btheta)$. There is some ambiguity in the definition of $\pW$ for binomial data depending on whether $Y_i$ is considered a unitary observation or is decomposed into $m_i$ Bernoulli random variables. Both models generate the same likelihood, but the perturbation space of the Bernoulli model is much larger. This does not affect the definitions of $\pD$ or $\pV$ but it does affect $\pW$. If $Y_i$ is considered the smallest unit of observation, then its contribution to $\pW$ is
\begin{equation}
\label{eq:pW_binomial}
p^{\text{bin}}_{Wi} = \text{Var}\left[ y_i \log (\pi_i) + (m_i - y_i) \log(1 - \pi_i)\right]
\end{equation}
If $Y_i$ is considered as the sum of $m_i$ independent Bernoulli trials, each of which can be individually perturbed, then their aggregate contribution to $\pW$ is
\begin{equation}
\label{eq:pW_bernoulli}
p^{\text{bern}}_{Wi} = y_i \text{Var}\left[\log (\pi_i)\right] +
(m_i - y_i) \text{Var}\left[\log(1 - \pi_i)\right]
\end{equation}
The appropriate expression to use depends on the context. If a binomial outcome is obtained by repeated Bernoulli trials on the same observational unit then $p^{\text{bin}}_{Wi}$ is more appropriate. Conversely, if a binomial random variable is created by post hoc aggregation of Bernoulli outcomes on different observational units that happen to share the same predictor variables then $p^{\text{bern}}_{Wi}$ is more appropriate.

\subsubsection{UNOS data}

The first example is adapted from \citet{Gelfand2003} using data from the United Network for Organ Sharing (UNOS). The data concerns $n=235$ heart transplants that took place in 10 centers across 5 age groups. Let $Y_{ca}$ be the number of adverse outcomes (organ rejection or death) that occurred out of $m_{ca}$ transplants in center $c$  and age group $a$. This is modeled by a mixed-effects logistic regression model
\begin{equation}
\label{eq:unos}
\begin{split}
Y_{ca} & \sim  \text{Bin}(\pi_{ca}, m_{ca}) \\
\log\left(\frac{\pi_{ca}}{1 - \pi_{ca}}\right) & = \mu_\alpha + \alpha_c + (\mu_\beta + \beta_c) X_a
\end{split}
\end{equation}
where $X_a$ is the midpoint age in group $a$, centered and scaled so that a unit increase in $X_a$ represents a 10-year age difference.

The parameters $\mu_\alpha, \mu_\beta$ are fixed effects and $\alpha_c, \beta_c$ are center-specific random effects for the intercept and slope.
\begin{align*}
\alpha_c & \sim N(0, \sigma^2_\alpha) \\
\beta_c  & \sim N(0, \sigma^2_\beta)
\end{align*}
with a sum-to-zero constraint $\sum_c \alpha_c = \sum_c \beta_c = 0$.

\begin{figure}[tbp]
\resizebox{\textwidth}{!}{\includegraphics{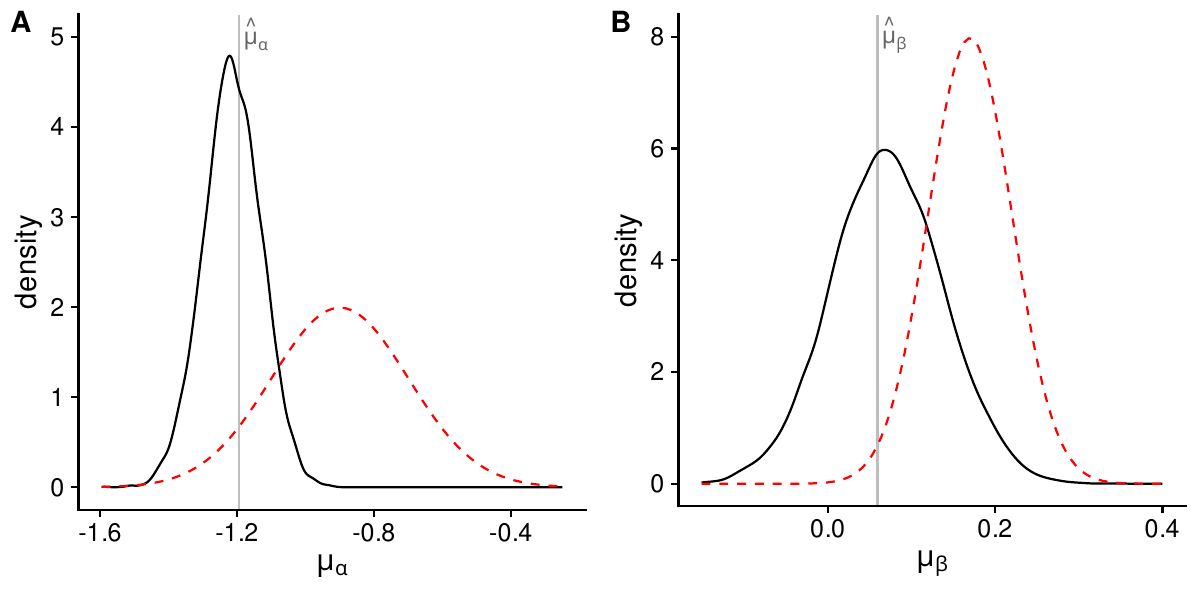}}
\caption{\label{fig:unos1} Solid curve: posterior density for fixed-effect parameters $\mu_\alpha$ (panel A) and $\mu_b$ (panel B) in the UNOS data model (\ref{eq:unos}) using a weakly informative prior. Vertical lines show the maximum likelihood estimates. Dotted curve: an alternative informative prior (\ref{eq:unos:informative})}
\end{figure}

Figure~\ref{fig:unos1} shows the posterior distributions of the fixed effects from a reference model that uses diffuse normal priors for $\mu_\alpha, \mu_\beta$ and weakly informative half-$t_2$ priors on $\sigma_\alpha, \sigma_\beta$. The posteriors from this model are consistent with maximum likelihood estimates given by the \texttt{glmer} function in the R package \textsf{lme4} \citep{bates2015}, shown by the vertical gray lines. Figure~\ref{fig:unos1} also shows an alternative informative prior represented by the dotted curves:
\begin{equation}
\label{eq:unos:informative}
\begin{split}
\mu_\alpha & \sim N(-0.9, 0.20^2) \\
\mu_\beta  & \sim N(0.17, 0.05^2)
\end{split}
\end{equation}
The informative prior puts prior probability mass away from  the maximum likelihood estimate. Focusing on $\mu_\alpha$, consider the effect of increasing the prior-data conflict in two ways: firstly by shifting the prior mean further away from $\widehat{\mu}_\alpha$, and secondly by concentrating the prior distribution about a fixed mean. The results, using the Bernoulli form of $\pW$ (\ref{eq:pW_bernoulli}), are shown in table~\ref{tab:unos1}.

\begin{table}[tbp]
\begin{center}
\begin{tabular}{rr|rrrc}
\hline
\multicolumn{2}{c|}{Prior on $\mu_\alpha$} & $\pD^*$ & $\pW$ & $\pV$ & $\pV/\pW$
\\
mean                 & sd                 \\
\hline
\multicolumn{2}{l|}{Reference} & 11.5 & 11.5 & 19.8 & 1.7 \\
\hline
\multicolumn{2}{l|}{Mean shift} & \\
-0.9 & 0.20 & 10.6 & 10.6 & 20.3 & 1.9\\
-0.7 & 0.20 & 10.3 & 10.2 & 21.6 & 2.1\\
-0.5 & 0.20 & 10.1 &  9.9 & 23.2 & 2.4\\
-0.3 & 0.20 &  9.9 &  9.5 & 24.9 & 2.6\\
-0.1 & 0.20 &  9.7 &  9.2 & 27.0 & 2.9\\
\hline
\multicolumn{2}{l|}{Concentration} & \\
-0.9 & 0.20 & 10.6 & 10.6 & 20.4 & 1.9\\
-0.9 & 0.10 &  9.7 &  9.4 & 22.0 & 2.3\\
-0.9 & 0.05 &  8.4 &  7.9 & 21.1 & 2.7\\
-0.9 & 0.02 &  7.6 &  7.0 & 15.8 & 2.3\\
-0.9 & 0.01 &  7.4 &  6.8 & 13.9 & 2.0\\
\hline
\end{tabular}
\caption{\label{tab:unos1} Effect of increasing prior-data conflict on penalty functions $(\pD^*), (\pW), (\pV)$ in the UNOS data. Mean shift moves the prior mean progressively away from the maximum likelihood estimate. Concentration progressively reduces the prior standard deviation. }
\end{center}
\end{table}

The first row of table~\ref{tab:unos1} shows the results for the reference model. The ratio $\pV/\pW=1.7$ may appear high for a prior distribution that is considered weakly informative. Note however that for non-informative priors we only expect $\pV \approx \pW$ asymptotically for large $n$. A probabilistic calibration of $\pV/\pW$ can be obtained using the posterior predictive testing framework of \citet{Gelman1996}. In this framework, replicate data sets are generated from the posterior predictive distribution of the data. The model is then re-fitted to each replicate data set to obtain a predictive distribution for the test statistic. For the UNOS data with the reference model, this gives a median value of $1.4$ for $\pV/\pW$ with a 95\% predictive interval of $(0.9, 1.9)$. By construction, none of the replicate data sets exhibit prior-data conflict. This shows the high variability of $\pV/\pW$ under the null hypothesis of no conflict and hence that the observed value of $1.7$ is not outstandingly large.

Row 2 of table~\ref{tab:unos1} shows results for the informative prior (\ref{eq:unos:informative}) which shows a modest increase in $\pV/\pW$ to 1.9. Rows 3-6 show the effect of shifting the prior mean further away from $\widehat{\mu}_\alpha$. Each shift, equal to 1 prior standard deviation, leads to a further increase in $\pV/\pW$.

Rows 7-11 of table~\ref{tab:unos1} show the effect of fixing the prior mean at $-0.9$ but reducing the prior standard deviation by approximately half each time. The ratio $\pV/\pW$ is not a monotonic function of the concentration. It reaches a maximum of 2.7 for a prior standard deviation sd=0.05, then decreases to 2.3 for sd=0.02 and 2.0 for sd=0.01. This may be explained by the fact that a sufficiently strong prior resolves prior-data conflict in favor of the prior, so that the posterior becomes insensitive to perturbations to the data. 

\subsubsection{Bristol Royal Infirmary inquiry data}

\begin{figure}[tbp]
\resizebox{\textwidth}{!}{\includegraphics{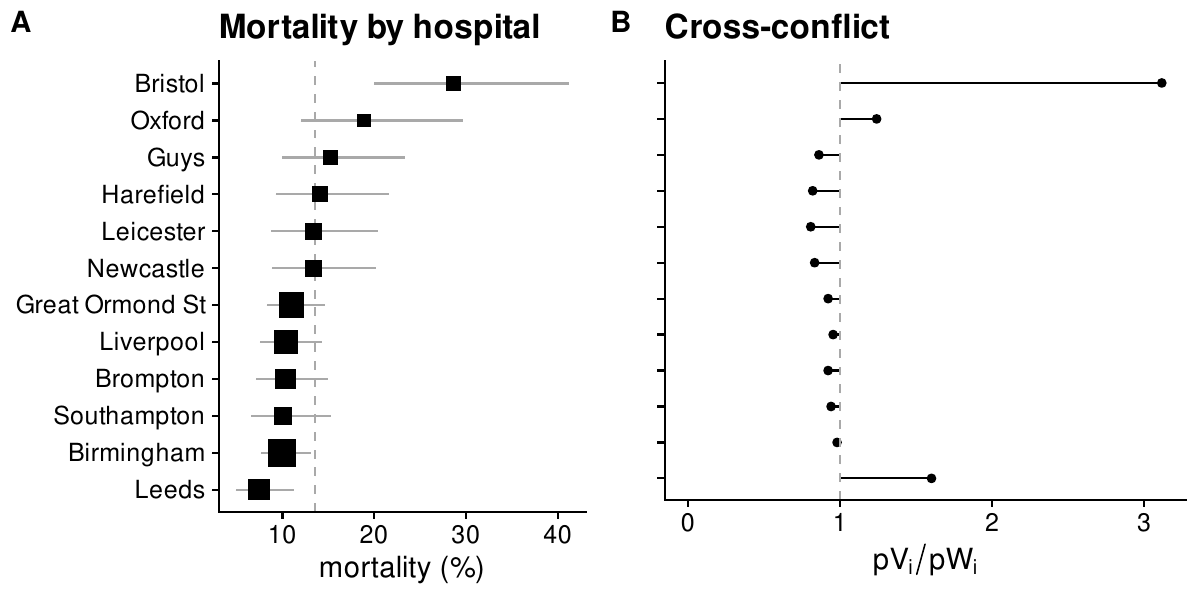}}
\caption{\label{fig:bri1} Data from the Bristol Royal Infirmary inquiry {\bf Panel A}: Estimated mortality and 95\% confidence intervals for 12 hospitals. 
The vertical dotted line shows the estimated population average mortality from a random-effects model (\ref{eq:bri_model}). {\bf Panel B}: Cross-conflict between hospitals}
\end{figure}

In addition to detecting global prior-data conflict, the ratio $\pV/\pW$ may be used to investigate divergent behavior at different levels of a hierarchical model. This is illustrated with data from the Bristol Royal Infirmary inquiry presented by \citet{Marshall2007}.  The inquiry investigated claims of excess mortality in pediatric cardiac surgery carried out at the Bristol Royal Infirmary prior to 1995. Panel A of figure~\ref{fig:bri1} shows mortality in 12 hospitals where comparable cardiac surgeries took place.

Let $Y_i$ be the number of deaths among $n_i$ operations carried out at hospital $i$. Then the data can be analyzed using a random effects logistic regression model
\begin{equation}
\label{eq:bri_model}
\begin{split}
Y_i & \sim \text{Bin}(\pi_i, n_i) \\
\log\left(\frac{\pi_i}{1 - \pi_i} \right) & = \theta_i \\
\theta_i & \sim N(\beta, \omega^2)
\end{split}
\end{equation}
with uniform priors on the hyper-parameters $\beta, \omega$.

Let $\Yvec_{-i}$ be the data for all hospitals except hospital $i$. If the focus of interest is on hospital $i$ then we can consider $p(Y_i \mid \theta_i)$ as the likelihood and $p(\theta_i \mid \bY_{-i})$ as a cross-validation prior distribution for $\theta_i$ that subsumes all information from the other hospitals. Then there are 12 sets of paired statistics $(p_{Wi}, p_{Vi})$, one for each hospital, where $p_{Wi}$ is defined by perturbing each of the $n_i$ operations individually as in equation (\ref{eq:pW_bernoulli}) and $p_{Vi}$ is defined by a single common perturbation to all operations in hospital $i$ as in equation (\ref{eq:pW_binomial}).  
Rather than measuring prior-data conflict, the ratio $p_{Vi}/p_{Wi}$ measures cross-conflict between different hospitals. Note that the ratio $p_{Vi}/p_{Wi}$ does not require an explicit calculation of of the cross-validation prior, nor does it require that $p(\theta_i \mid \Yvec_{-i})$ is expressed in closed form.

The results are shown in panel B of figure~\ref{fig:bri1}. As might be expected from the context, the Bristol Royal Infirmary shows the largest cross-conflict with $p_{Vi}/p_{Wi} = 3.1$. This is roughly double the cross-conflict for Leeds $(p_{Vi}/p_{Wi} = 1.6)$, where mortality was lower than average (Panel A). The remaining hospitals show cross-conflict values close to 1.

\section{Discussion}
\label{section:discussion}

Bayesian sensitivity analysis has often concentrated on prior sensitivity \citep{Roos2015} since the presence of a prior distribution distinguishes Bayesian from frequentist inference. Bayesian and frequentist models share a common problem that they may be sensitive to changes in the data. The sensitivity diagnostics discussed here do not test any specific deviation from the working model, and hence may point to a variety of different issues, including: errors in the data, failure to capture certain data features in the working model, and conflict between the data and an informative prior.

There is a strong connection between local influence and predictive information criteria. All of these criteria are based on the same principle that optimism in a numeric assessment of model fit can be offset by adding a complexity penalty. Over-fitted models are likely to be both highly optimistic -- hence requiring a large penalty -- and be sensitive to small changes to the data. So intuitively it is not surprising to find this connection. Nevertheless, it is interesting that the penalties used by WAIC and DIC can be derived from the same local sensitivity framework based on phi-divergences. This connection may also help to understand the differences between WAIC and DIC. The WAIC penalty $\pW$ is an aggregate measure of influence whereas the alternative DIC penalty $\pD^*$ is an aggregate measure of leverage. This suggests that WAIC will be more sensitive to highly influential observations. The Bayesian leverage $LLEV$ is derived from a posterior predictive framework and is closely related to Value of Information analysis, in which decisions are guided by the assumption that the model is true. Hence DIC incorporates an assumption that the model is correct.

The ratio $\pV/\pW$ measures sensitivity to perturbations of the learning rate that may reveal prior-data conflict. While $\pW$ and $\pD^*$ are often numerically very similar, section~\ref{section:prior_data_conflict} shows that $\pV$ may penalize models much more strongly in the presence of prior-data conflict. This phenomenon might actually make $\pV$ a more attractive penalty for model choice since we would normally want to choose a model without prior-data conflict. However, $\pV$ currently only has a heuristic justification in the model choice context. The $\pV/\pW$ ratio may be re-purposed to diagnose cross-conflict between different parts of the data. This may have interesting applications in modular inference, where the presence of cross-conflict is the justification for departing from the standard rules of Bayesian updating \citep{Carmona2020}.


\section*{Data availability}

The Abalone data (\url{https://doi.org/10.24432/C55C7W}) and Bike Sharing data (\url{https://doi.org/10.24432/C5W894}) are both available from the UCI machine learning repository. The UNOS data were tabulated by \citet{Gelfand2003} and the Bristol Royal Infirmary data by \citet{Marshall2007}.

\if1\anon{
\appendix
\section{Supplementary Material}

R and JAGS code for all examples is available at

\url{https://github.com/martynplummer/bayesian-influence}. 

\noindent
The examples require JAGS 5.0.0 or later.
} \fi

\clearpage

\bibliographystyle{apalike}

\begingroup
    \setlength{\bibsep}{12pt}
    \spacingset{1}
    \bibliography{influence}

@article{Watanabe2010,
  author  = {Sumio Watanabe},
  title   = {Asymptotic Equivalence of Bayes Cross Validation and Widely Applicable Information Criterion in Singular Learning Theory},
  journal = {Journal of Machine Learning Research},
  year    = {2010},
  volume  = {11},
  number  = {116},
  pages   = {3571--3594},
  url     = {http://jmlr.org/papers/v11/watanabe10a.html}
}

@book{Watanabe2009Book,
  title={Algebraic Geometry and Statistical Learning Theory},
  author={Watanabe, Sumio},
  year={2009},
  publisher={Cambridge University Press}
}

@article{Spiegelhalter2002DIC,
  title={Bayesian Measures of Model Complexity and Fit},
  author={Spiegelhalter, David J. and Best, Nicola G. and Carlin, Bradley P. and van der Linde, Angelika},
  journal={Journal of the Royal Statistical Society: Series B (Statistical Methodology)},
  volume={64},
  number={4},
  pages={583--639},
  year={2002},
  publisher={Wiley Online Library}
}

@article{Spiegelhalter2014,
  title={The deviance information criterion: 12 years on},
  author={Spiegelhalter, David J. and Best, Nicola G. and Carlin, Bradley P. and van der Linde, Angelika},
  journal={Journal of the Royal Statistical Society: Series B (Statistical Methodology)},
  volume={76},
  number={3},
  pages={485--493},
  year={2014},
  publisher={Wiley Online Library}
}

@book{Gelman2004BDA,
  title={Bayesian Data Analysis},
  author={Gelman, Andrew and Carlin, John B. and Stern, Hal S. Rubin, Donald B.},
  year={2004},
  edition={2},
  publisher={Chapman and Hall/CRC}
}

@article{MillarStewart2007,
author = {Russell B. Millar and Wayne S. Stewart},
title = {{Assessment of locally influential observations in {B}ayesian models}},
volume = {2},
journal = {Bayesian Analysis},
number = {2},
publisher = {International Society for Bayesian Analysis},
pages = {365 -- 383},
year = {2007},
doi = {10.1214/07-BA216},
URL = {https://doi.org/10.1214/07-BA216}
}

@article{vanderLinde2007,
author = {Angelika van der Linde},
title = {{Local influence on posterior distributions under multiplicative modes of perturbation}},
volume = {2},
journal = {Bayesian Analysis},
number = {2},
publisher = {International Society for Bayesian Analysis},
pages = {319 -- 332},
year = {2007},
doi = {10.1214/07-BA213},
URL = {https://doi.org/10.1214/07-BA213}
}

@article{DrtonPlummer2017,
author={Drton,Mathias and Plummer,Martyn},
year={2017},
title={A {B}ayesian information criterion for singular models},
journal={Journal of the Royal Statistical Society. Series B, Statistical methodology},
volume={79},
number={2},
pages={323-380},
isbn={1369-7412},
language={English},
}

@article{Gelman2014,
author={Gelman,Andrew and Hwang,Jessica and Vehtari,Aki},
year={2014},
title={Understanding predictive information criteria for {B}ayesian models},
journal={Statistics and computing},
volume={24},
number={6},
pages={997-1016},
isbn={0960-3174},
language={English},
}

@article{Akaike1973,
  title={Information Theory and an Extension of the Maximum Likelihood Principle},
  author={Akaike, Hirotugu},
  journal={Proceedings of the Second International Symposium on Information Theory},
  year={1973},
  pages={267--281}
}

@article{Brooks2002discussion,
  title={Discussion on the paper by Spiegelhalter, Best, Carlin and van der Linde},
  author={Brooks, S. P. and Smith, J. and Vehtari, A. and Plummer, M. and Stone, M. and Robert, C. P. and Titterington, D. M. and Nelder, J. A. and Atkinson, A. and Dawid, A. P. and others},
  journal={Journal of the Royal Statistical Society: Series B (Statistical Methodology)},
  volume={64},
  number={4},
  pages={616--639},
  year={2002}
}

@misc{Abalone94,
  author = {Nash, Warwick and Sellers, Tracy and Talbot, Simon and Cawthorn, Andrew and Ford, Wes},
  title = {{Abalone}},
  year = {1994},
  howpublished = {{UCI} Machine Learning Repository},
  doi = {10.24432/C55C7W},
  URL = {https://doi.org/10.24432/C55C7W}
}

@misc{BikeSharing2013,
  author       = {Fanaee-T, Hadi},
  title        = {{Bike Sharing}},
  year         = {2013},
  howpublished = {{UCI} Machine Learning Repository},
  doi          = {10.24432/C5W894},
  URL          = {https://doi.org/10.24432/C5W894},
}

@incollection{Gelfand2003,
  title={Some comments on model criticism},
  author={Gelfand, Alan E.},
  booktitle={Highly Structured Stochastic Systems},
  editor={Green, Peter J. and Hjort, Nils Lid and Richardson, Sylvia},
  pages={449--454},
  year={2003},
  publisher={Oxford University Press},
  address={Oxford, UK}
}

@Article{Bates2015,
    title = {Fitting Linear Mixed-Effects Models Using {lme4}},
    author = {Douglas Bates and Martin M{\"a}chler and Ben Bolker and Steve Walker},
    journal = {Journal of Statistical Software},
    year = {2015},
    volume = {67},
    number = {1},
    pages = {1--48},
    doi = {10.18637/jss.v067.i01},
}

@article{Marshall2007,
    author={Marshall,E. C. and Spiegelhalter,D. J.},
    year={2007},
    title={Identifying outliers in {B}ayesian hierarchical models: a simulation-based approach},
    journal={Bayesian Analysis},
    volume={2},
    number={2},
    isbn={1936-0975},
    language={English}
}

@article{Thomas2018,
  author={Thomas,Zachary M. and MacEachern,Steven N. and Peruggia,Mario},
  year={2018},
  title={Reconciling Curvature and Importance Sampling Based Procedures for Summarizing Case Influence in {B}ayesian Models},
  journal={Journal of the American Statistical Association},
  volume={113},
  number={524},
  pages={1669-1683},
  isbn={0162-1459},
  language={English},
}

@article{Jackson2022,
  author={Jackson, Christopher H. and Baio, Gianluca and Heath, Anna and Strong, Mark and Welton, Nicky J. and Wilson, Edward C. F.},
  year={2022},
  title={Value of Information Analysis in Models to Inform Health Policy},
  journal={Annual review of statistics and its application},
  volume={9},
  number={1},
  pages={95-118},
  isbn={2326-8298},
  language={English},
}

@article{Cook1986,
  author={Cook,R. D.},
  year={1986},
  title={Assessment of Local Influence},
  journal={Journal of the Royal Statistical Society. Series B, Methodological},
  volume={48},
  number={2},
  pages={133-169},
  isbn={0035-9246},
  language={English},
}

@article{McCulloch1989,
  author={McCulloch,Robert E.},
  year={1989},
  title={Local Model Influence},
  journal={Journal of the American Statistical Association},
  volume={84},
  number={406},
  pages={473-478},
  isbn={0162-1459},
  language={English},
}

@article{Lavine1992,
  author={Lavine, Michael},
  year={1992},
  title={Local predictive influence in {B}ayesian linear models with conjugate priors},
  journal={Communications in statistics. Simulation and computation},
  volume={21},
  number={1},
  pages={269-283},
  isbn={0361-0918},
  language={English},
}

@article{Poon1999,
  author={Poon, W.-Y. and Poon,Y. S.},
  year={1999},
  title={Conformal normal curvature and assessment of local influence},
  journal={Journal of the Royal Statistical Society. Series B, Statistical methodology},
  volume={61},
  number={1},
  pages={51-61},
  isbn={1369-7412},
  language={English},
}

@article{ZhuLee2001,
  author={Zhu,Hong-Tu and Lee,Sik-Yum},
  year={2001},
  title={Local influence for incomplete data models},
  journal={Journal of the Royal Statistical Society. Series B, Statistical methodology},
  volume={63},
  number={1},
  pages={111-126},
  isbn={1369-7412},
  language={English},
}

@article{Millar2018,
  Author = {Millar, Russell B.},
  Title = {Conditional vs marginal estimation of the predictive loss of
   hierarchical models using WAIC and cross-validation},
  Journal = {STATISTICS AND COMPUTING},
  Year = {2018},
  Volume = {28},
  Number = {2},
  Pages = {375-385},
  Month = {MAR},
  DOI = {10.1007/s11222-017-9736-8},
  ISSN = {0960-3174},
  EISSN = {1573-1375},
  Unique-ID = {WOS:000422798100009},
}

@article{Parsons2022,
   author={Parsons, Jacob and Bao, Le},
   year={2022},
   title={A unified approach for outliers and influential data detection: The value of information in retrospect},
   journal={Stat (International Statistical Institute)},
   volume={11},
   number={1},
   pages={e442},
   isbn={2049-1573},
   language={English},
}

@article{Demidenko2005,
  author={Demidenko,Eugene and Stukel,Therese A.},
  year={2005},
  title={Influence analysis for linear mixed-effects models},
  journal={Statistics in medicine},
  volume={24},
  number={6},
  pages={893-909},
  isbn={0277-6715},
  language={English},
}

@article{Nobre2011,
  author={Nobre,Juvêncio S. and Singer,Julio M.},
  year={2011},
  title={Leverage analysis for linear mixed models},
  journal={Journal of applied statistics},
  volume={38},
  number={5},
  pages={1063-1072},
  isbn={0266-4763},
  language={English},
}

@article{Christensen1992,
  author = {Ronald Christensen and Larry M. Pearson and Wesley Johnson},
  title = {Case-Deletion Diagnostics for Mixed Models},
  journal = {Technometrics},
  volume = {34},
  number = {1},
  pages = {38--45},
  year = {1992},
  publisher = {ASA Website},
  doi = {10.1080/00401706.1992.10485231},
  URL = {https://www.tandfonline.com/doi/abs/10.1080/00401706.1992.10485231},
}

@article{Lesaffre1998,
  author={Lesaffre,Emmanuel and Verbeke,Geert},
  year={1998},
  title={Local Influence in Linear Mixed Models},
  journal={Biometrics},
  volume={54},
  number={2},
  pages={570-582},
  isbn={0006-341X},
  language={English},
}

@article{Beckman1987,
  author={Beckman,Richard J. and Nachtsheim,Christopher J. and Cook,R. D.},
  year={1987},
  title={Diagnostics for Mixed-Model Analysis of Variance},
  journal={Technometrics},
  volume={29},
  number={4},
  pages={413},
  isbn={0040-1706},
  language={English},
}

@article{Ouwens2001,
  author={Ouwens, Mario J. N. M and Tan,Frans E. S. and Berger,Martijn P. F.},
  year={2001},
  title={Local Influence to Detect Influential Data Structures for Generalized Linear Mixed Models},
  journal={Biometrics},
  volume={57},
  number={4},
  pages={1166-1172},
  isbn={0006-341X},
  language={English},
}

@article{Rakhmawati2017,
  author = {Trias Wahyuni Rakhmawati and Geert Molenberghs and Geert Verbeke and Christel Faes},
  title = {Local influence diagnostics for generalized linear mixed models with overdispersion},
  journal = {Journal of Applied Statistics},
  volume = {44},
  number = {4},
  pages = {620--641},
  year = {2017},
  publisher = {Taylor \& Francis},
  doi = {10.1080/02664763.2016.1182128},
  URL = {https://doi.org/10.1080/02664763.2016.1182128}
}

@book{CookWeisberg1982,
  author={Cook,R. D. and Weisberg, S.},
  year={1982},
  title={Residuals and influence in regression},
  publisher={Chapman and Hall},
  address={London;New York;},
  isbn={9780412242809;041224280X;},
  language={English},
}

@article{Roos2015,
  author={Roos,Małgorzata and Martins,Thiago G. and Held,Leonhard and Rue,Håvard},
  year={2015},
  title={Sensitivity Analysis for {B}ayesian Hierarchical Models},
  journal={Bayesian Analysis},
  volume={10},
  number={2},
  isbn={1936-0975},
  language={English},
}

@InProceedings{Carmona2020,
  title = 	 {Semi-Modular Inference: enhanced learning in multi-modular models by tempering the influence of components},
  author =       {Carmona, Christian and Nicholls, Geoff},
  booktitle = 	 {Proceedings of the Twenty Third International Conference on Artificial Intelligence and Statistics},
  pages = 	 {4226--4235},
  year = 	 {2020},
  editor = 	 {Chiappa, Silvia and Calandra, Roberto},
  volume = 	 {108},
  series = 	 {Proceedings of Machine Learning Research},
  month = 	 {26--28 Aug},
  publisher =    {PMLR},
  url = 	 {https://proceedings.mlr.press/v108/carmona20a.html},
}

@article{Yang2024,
   title={Generalized Out-of-Distribution Detection: A Survey},
   author={Yang, Jingkang and Zhou, Kaiyang and Li, Yixuan and Liu, Ziwei},
   journal={International Journal of Computer Vision},
   year={2024},
   pages={5635--5662},
   volume={132},
   url={UR  - https://doi.org/10.1007/s11263-024-02117-4},
   doi={10.1007/s11263-024-02117-4}
}

@article{Salehi2022,
  title={A Unified Survey on Anomaly, Novelty, Open-Set, and Out of-Distribution Detection: Solutions and Future Challenges},
  author={Mohammadreza Salehi and Hossein Mirzaei and Dan Hendrycks and Yixuan Li and Mohammad Hossein Rohban and Mohammad Sabokrou},
  journal={Transactions of Machine Learning Research},
  year={2022},
  url={https://openreview.net/forum?id=aRtjVZvbpK},
}

@article{Watanabe2013,
  author  = {Sumio Watanabe},
  title   = {A Widely Applicable {B}ayesian Information Criterion},
  journal = {Journal of Machine Learning Research},
  year    = {2013},
  volume  = {14},
  number  = {27},
  pages   = {867--897},
  url     = {http://jmlr.org/papers/v14/watanabe13a.html}
}

@article{MillerDunson2019,
 ISSN = {01621459},
 URL = {http://www.jstor.org/stable/45218583},
 author = {Jeffrey W. Miller and David B. Dunson},
 journal = {Journal of the American Statistical Association},
 number = {527},
 pages = {1113--1125},
 publisher = {[American Statistical Association, Taylor & Francis, Ltd.]},
 title = {Robust {B}ayesian Inference via Coarsening},
 urldate = {2025-11-11},
 volume = {114},
 year = {2019}
}

@article{Bhattacharya2019,
  author={Bhattacharya,Anirban and Pati,Debdeep and Yang,Yun},
  year={2019},
  title={Bayesian Fractional Posteriors},
  journal={The Annals of statistics},
  volume={47},
  number={1},
  pages={39-66},
  isbn={0090-5364},
  language={English},
}

@article{GrunwaldVanOmmen2017,
  author={Grünwald,Peter and van Ommen,Thijs},
  year={2017},
  title={Inconsistency of {B}ayesian Inference for Misspecified Linear Models, and a Proposal for Repairing It},
  journal={Bayesian Analysis},
  volume={12},
  number={4},
  isbn={1936-0975},
  language={English},
}

@article{Gelman1996,
 ISSN = {10170405, 19968507},
 URL = {http://www.jstor.org/stable/24306036},
 author = {Andrew Gelman and Xiao-Li Meng and Hal Stern},
 journal = {Statistica Sinica},
 number = {4},
 pages = {733--760},
 publisher = {Institute of Statistical Science, Academia Sinica},
 title = {POSTERIOR PREDICTIVE ASSESSMENT OF MODEL FITNESS VIA REALIZED DISCREPANCIES},
 urldate = {2025-11-24},
 volume = {6},
 year = {1996}
}

@article{Plummer2008,
    author = {Plummer, Martyn},
    title = {Penalized loss functions for {B}ayesian model comparison},
    journal = {Biostatistics},
    volume = {9},
    number = {3},
    pages = {523-539},
    year = {2008},
    month = {01},
    issn = {1465-4644},
    doi = {10.1093/biostatistics/kxm049},
    url = {https://doi.org/10.1093/biostatistics/kxm049},
    eprint = {https://academic.oup.com/biostatistics/article-pdf/9/3/523/17742541/kxm049.pdf},
}
\endgroup

\end{document}